\documentclass[%
 aip,superscriptaddress
 amsmath,amssymb,
 reprint,%
]{revtex4-1}

\usepackage[colorlinks,
linkcolor=blue, urlcolor=blue, anchorcolor=blue, citecolor=blue]{hyperref}
\usepackage{graphicx}
\usepackage{dcolumn}
\usepackage{bm}

\usepackage[utf8]{inputenc}
\usepackage[T1]{fontenc}
\usepackage{mathptmx}
\usepackage{booktabs}
\usepackage{cleveref}
\usepackage{subfigure}

\begin{document}

\preprint{AIP/123-QED}

\title[Experimental Analysis of PandaX-4T Cryogenic Distillation System for Removing Krypton from Xenon]{Experimental Analysis of PandaX-4T Cryogenic Distillation System for Removing Krypton from Xenon\\
\vbox{}}

\author{Rui Yan}
\affiliation{ Institute of Refrigeration and Cryogenics, Shanghai Jiao Tong University, Shanghai, 200240, China}

\author{Zhou Wang}
\altaffiliation{Author to whom correspondence should be addressed:\\wangzhou0303@sjtu.edu.cn\\yju@sjtu.edu.cn}
\affiliation{ Institute of Particle and Nuclear Physics, Shanghai Jiao Tong University, Shanghai, 200240, China}
\affiliation{ Shanghai Jiao Tong University Sichuan Research Institute, Chengdu, 610000, China}

\author{Xiangyi Cui}
\affiliation{ Shanghai Jiao Tong University Sichuan Research Institute, Chengdu, 610000, China}
\affiliation{Tsung-Dao Lee Institute, Shanghai Jiao Tong University, Shanghai, 200240, China}

\author{Yonglin Ju}
\altaffiliation{Author to whom correspondence should be addressed:\\wangzhou0303@sjtu.edu.cn\\yju@sjtu.edu.cn}
\affiliation{ Institute of Refrigeration and Cryogenics, Shanghai Jiao Tong University, Shanghai, 200240, China}

\author{Haidong Sha}
\affiliation{ Institute of Refrigeration and Cryogenics, Shanghai Jiao Tong University, Shanghai, 200240, China}

\author{Shuaijie Li}
\affiliation{Tsung-Dao Lee Institute, Shanghai Jiao Tong University, Shanghai, 200240, China}

\author{Peiyao Huang}
\affiliation{ Institute of Refrigeration and Cryogenics, Shanghai Jiao Tong University, Shanghai, 200240, China}

\author{Xiuli Wang}
\affiliation{ Institute of Refrigeration and Cryogenics, Shanghai Jiao Tong University, Shanghai, 200240, China}

\author{Wenbo Ma}
\affiliation{ Institute of Particle and Nuclear Physics, Shanghai Jiao Tong University, Shanghai, 200240, China}

\author{Yingjie Fan}
\affiliation{School of Physics, Nankai University, Tianjin 300071, China}

\author{Li Zhao}
\affiliation{ Institute of Particle and Nuclear Physics, Shanghai Jiao Tong University, Shanghai, 200240, China}
\affiliation{ Shanghai Jiao Tong University Sichuan Research Institute, Chengdu, 610000, China}

\author{Jianglai Liu}
\affiliation{ Institute of Particle and Nuclear Physics, Shanghai Jiao Tong University, Shanghai, 200240, China}
\affiliation{ Shanghai Jiao Tong University Sichuan Research Institute, Chengdu, 610000, China}

\author{Xiangdong Ji}
\affiliation{ Institute of Particle and Nuclear Physics, Shanghai Jiao Tong University, Shanghai, 200240, China}
\affiliation{ Shanghai Jiao Tong University Sichuan Research Institute, Chengdu, 610000, China}
\affiliation{Tsung-Dao Lee Institute, Shanghai Jiao Tong University, Shanghai, 200240, China}
\affiliation{Department of Physics, University of Maryland, College Park, Maryland, 20742, USA}

\author{Jifang Zhou}
\affiliation{Yalong River Hydropower Development Company, Ltd.,
288 Shuanglin Road, Chengdu 610051, China}

\author{Changsong Shang}
\affiliation{Yalong River Hydropower Development Company, Ltd.,
288 Shuanglin Road, Chengdu 610051, China}

\author{Liqiang Liu}
\affiliation{Yalong River Hydropower Development Company, Ltd.,
288 Shuanglin Road, Chengdu 610051, China}

\date{\today}

\begin{abstract}
An efficient cryogenic distillation system was designed and constructed for PandaX-4T dark matter detector based on the McCabe-Thiele (M-T) method and the conservation of mass and energy. This distillation system is designed to reduce the concentration of krypton in commercial xenon from 5$\times 10^{-7}$ mol/mol to $\sim 10^{-14}$ mol/mol with 99 $\%$ xenon collection efficiency at a maximum flow rate of 10 kg/h. The offline distillation operation has been completed and 5.75 tons of ultra-high purity xenon was produced, which is used as the detection medium in PandaX-4T detector. The krypton concentration of the product xenon is measured with an upper limit of 8.0 ppt. The stability and purification performance of the cryogenic distillation system are studied by analyzing the experimental data, which is important for theoretical research and distillation operation optimization.
\end{abstract}

\maketitle

\section{Introduction\label{sec1}}

Dark matter particle detection is one of the most cutting-edge research topics in the field of astroparticle and particle physics. So far, the Weakly Interacting Massive Particles (WIMPs) is the most favorite candidate in terms of dark matter particle.\cite{1,2,3,4} Xenon (Xe) has been regarded as one of the most attractive medium for dark matter detectors because of the following advantages: (1) xenon can shield $\alpha$-rays and $\beta$-rays from uranium and thorium, (2) low energy threshold and high energy resolution, (3) high ionization yield and flicker light yield, (4) no long-lived radioactive isotope, and the background of xenon itself is very low.\cite{5,6,7,8}

Krypton-85 ($^{85}$Kr) is a radioactive nucleus which emits $\beta$-rays and thus it interferes with the detection of dark matter signals. To meet the strict background requirements of the dark matter detection experiment, the level of $^{85}$Kr should be decreased to at least ~$10^{-23}$ mol/mol. For the natural abundance of $^{85}$Kr accounts for $^{85}$Kr/Kr = $10^{-11}$ mol/mol in krypton,\cite{9} the ratio of krypton to xenon is required to be lower than $10^{-12}$ mol/mol (1 part per trillion, 1 ppt ). However, the commercial xenon separated from the atmosphere contains 5$\times 10^{-7}$ mol/mol (0.5 parts per million, 0.5 ppm ) of krypton. Therefore, it is critical to purify the commercial xenon in order to obtain ppt-level xenon for using in high-sensitivity and low-background dark matter detection experiments. 

One mature technique to achieve the ultra-pure xenon is cryogenic distillation which has been proven effective in several experiments. The XMASS collaboration applied cryogenic distillation to produce ultra-high purity xenon for dark matter detectors firstly, and reached a reduction of krypton level of 3 orders of magnitudes with the flow rate of 0.6 kg/h for one pass.\cite{9} Utilizing the distillation technique, PandaX-I and PandaX-II also obtained 3 orders of magnitudes of krypton reduction in xenon while increasing the distillation flow rate up to 5 kg/h for one pass.\cite{10,11} The upgraded distillation system of XMASS could reduce the amount of krypton by five orders of magnitude with one pass at a process flow rate of 4.7 kg/h.\cite{12} The krypton reduction factor of 6.5$\times 10^{5}$ was reached recently using the distillation column developed by Xenon1T with the flow rate of 3 kg/h.\cite{13,14}

Considering the weak interaction between dark matter and ordinary matter, PandaX collaboration group is committed to improving the sensitivity of dark matter detector continuously in order to to find the nature of dark matter. There are two effective methods to develop a more sensitive detector: one is to produce the higher purity xenon, the other is to increase the mass of liquid xenon in detector. The target liquid xenon of the next generation of dark matter detector called PandaX-4T is 4 tons, and the dark matter research plans to conduct under the krypton background at sub-ppt-level (~$10^{-13}$ mol/mol).\cite{15} Because the distillation system developed for PandaX-I and PandaX-II cannot reach the required xenon purity of sub-ppt krypton concentration and the processing rate is not fast enough for the PandaX-4T detector, a new cryogenic distillation system was designed and constructed to reduce the concentration of krypton in commercial xenon from 5$\times 10^{-7}$ mol/mol to $\sim 10^{-14}$ mol/mol with 99$\%$ xenon collection efficiency at the maximum flow rate of 10 kg/h. Under the design conditions, the newly developed cryogenic distillation system for PandaX-4T can be operated offline independently as well as online to be coupled with the dark matter detector for removal of krypton.\cite{16,17}

The PandaX-4T distillation system has worked independently for 1.5 months without coupling with the detector and purified 5.75 tons of xenon for using in the dark matter detector. In this paper, the main parameters of the cryogenic distillation system are briefly introduced in Section~\ref{sec2}, the operation modes are discussed in Section~\ref{sec3}, the experimental data analysis results of the experimental data are shown in Section~\ref{sec4}, and the krypton concentration of distillated product xenon is discussed in Section~\ref{sec5}, before we conclude in Section~\ref{sec6}.

\section{Main parameters of the distillation system\label{sec2}}

The McCabe-Thiele (M-T) method\cite{18} is adopted to calculate the number of the theoretical plates for the PandaX-4T distillation tower. The main design specifications of PandaX-4T distillation column are given as follows:
\begin{enumerate}
\item [1)] The concentration of krypton in commercial xenon should be reduced by 7 orders of magnitude, which means Kr/Xe should be decreased from 0.5 ppm to 0.01 ppt.
\item [2)] The distillation system operates at the flow rate of 10 kg/h.
\item [3)] The xenon collection efficiency should be 99$\%$.
\item [4)] The commercial original xenon enters into the distillation system as saturated liquid phase, and the corresponding reflux ratio (R) is 45.
\item [5)]The working temperature of the cryogenic distillation system is 178K, and the ratio of volatility between Kr and Xe (~$\alpha ^{Kr}$) is 10.7
\end{enumerate}

\begin{figure}[htbp]
\centering 
\includegraphics[width=.45\textwidth]{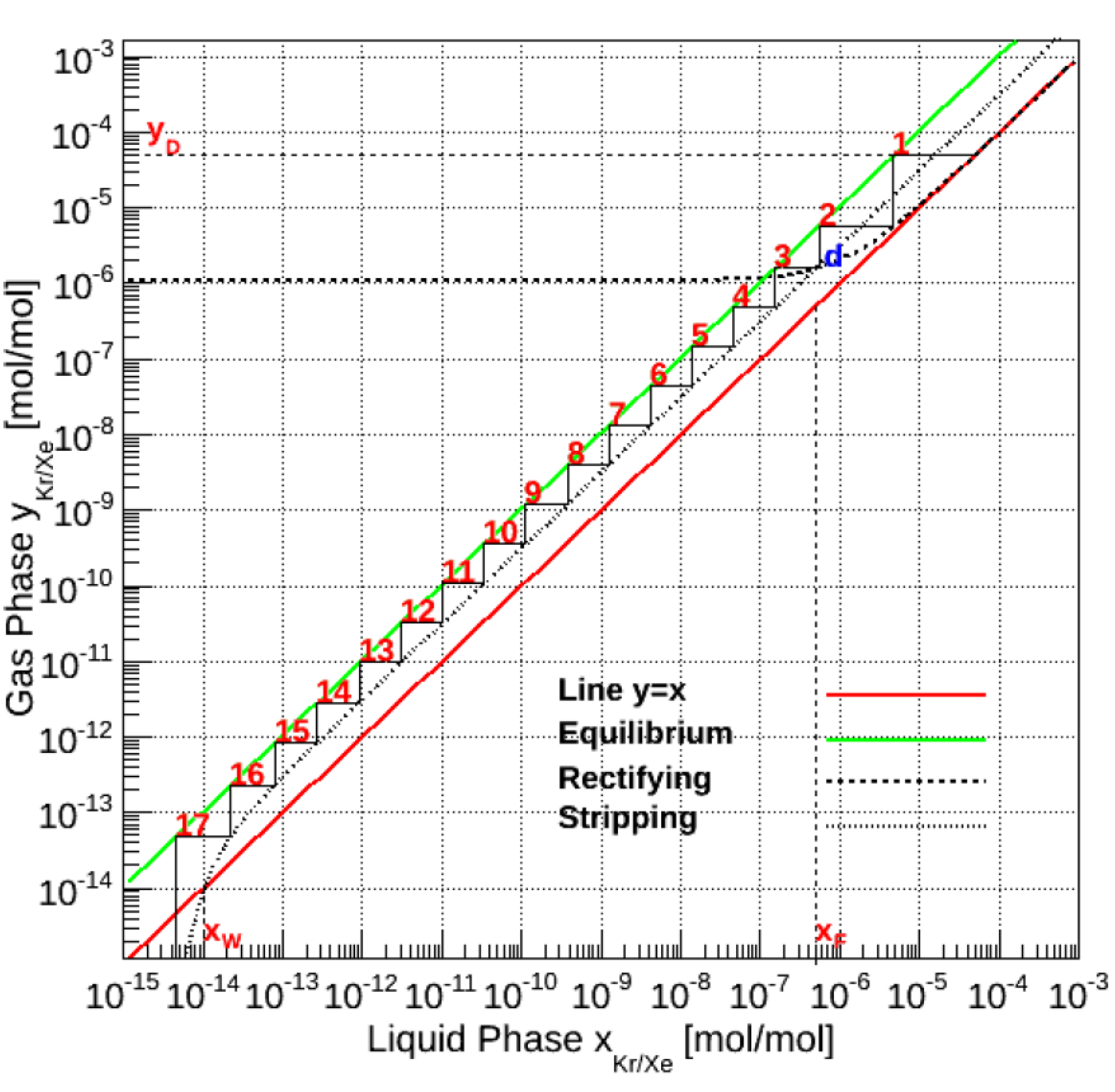}
\caption{M-T diagram under the condition of liquid feed with ~$\alpha ^{Kr}$ is 10.7 and reflux ratio is 45.\cite{17} \label{fig1}}
\end{figure}

Based on the design requirements listed above, the corresponding gas-liquid equilibrium curve, rectifying line and stripping line are drawn on M-T diagram, as shown in Fig.~\ref{fig1}. It can be seen from Fig.\ref{fig1} that the distillation tower can produce the ultra-high purity xenon as aimed (0.01 ppt) when the number of the theoretical plates (n) is 17. According to the calculation results, the feeding point should be at the third theoretical plate. The height equivalent of a theoretical plate (HETP) is 35 cm, which is the same value used in the first generation of the PandaX distillation system\cite{10}. The height of the distillation column is 6 m which corresponds to H=n$\cdot$HETP, and the feeding point is 1.5 m counting from the top.

The detailed design calculations about the geometrical dimensions and cold and heat source loads of the distillation tower were described in Ref.~[16, 17]. The main design parameters of the distillation system are given in Table~\ref{table1}.

\begin{table}[htbp]
    \caption{Main parameters of the distillation system.\cite{17} \label{table1}}
    \begin{tabular}{ccc}
    \toprule
    Parameters&  &Value\\
    \midrule
    Structured packing type&  &PACK-$^{13}$C\cite{19}\\
    HETP&  &35 cm\\
    Height of the packing column&  &6 m\\
    Height of the rectifying section&  &1.5 m\\
    Height of the stripping section&  &4.5 m\\
    Diameter of the distillation column&  &125 mm\\
    Feeding plate&  &3th\\
    Pre-cooling capacity&  &57 W\\
    Cooling capacity&  &373 W\\
    Heating capacity of the reboiler&  &120 W\\
    \toprule
    \end{tabular}
\end{table}

\section{Operation of the distillation system\label{sec3}}
\subsection{The technological process}

\begin{figure*}[htbp]
\centering 
\includegraphics[width=\textwidth]{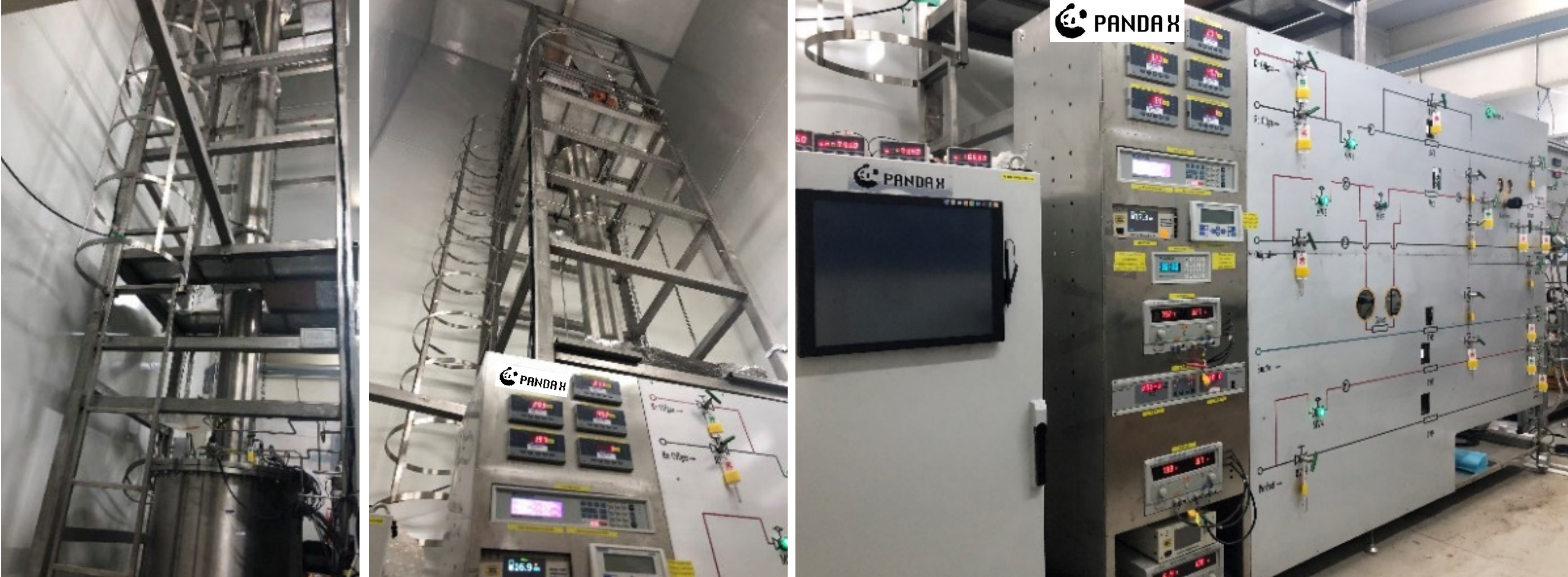}
\caption{Photos of the PandaX-4T distillation system.\label{fig2}}
\end{figure*}

The PandaX-4T distillation system was assembled in China Jinping Underground Laboratory (CJPL)\cite{20}. The photos of the constructed system can be seen in Fig.~\ref{fig2}. After several months of commissioning, the system passed the test run successfully. Then the offline operation of the krypton removal was implemented and the total amount of 5.75 tons of ultra-pure xenon was produced during 1.5 months for the PandaX-4T dark matter detector.

\begin{figure*}[htbp]
\centering 
\includegraphics[width=\textwidth]{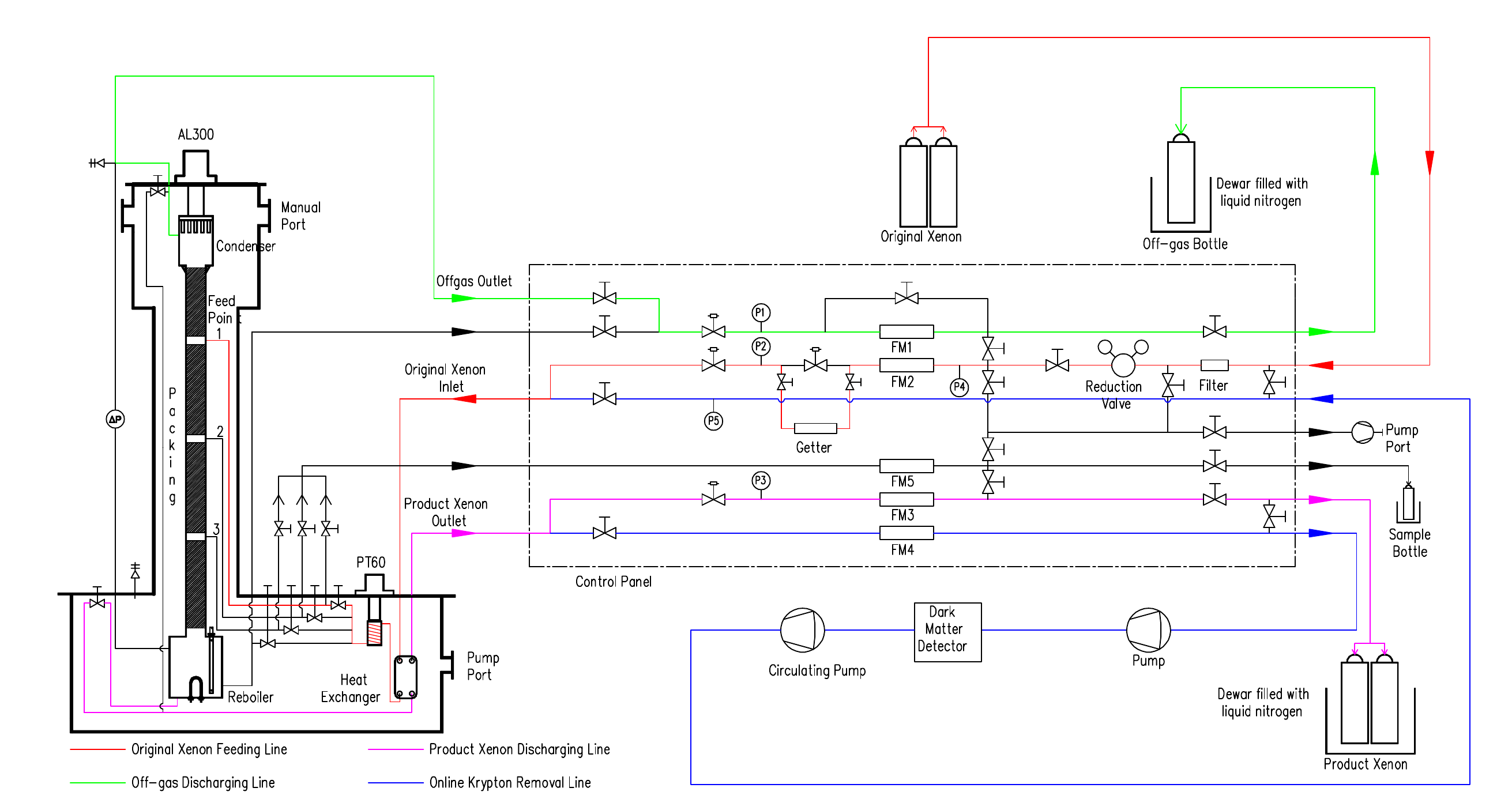}
\caption{The technological process of the distillation system.\cite{16} FM1-FM5: Flow controllers; AL300, PT60: Cryo-coolers. \label{fig3}}
\end{figure*}

There are three operating modes of PandaX-4T distillation system. One is the offline distillation mode in which the distillation system works independently for the removal of krypton. While the other two are online distillation modes in which the distillation system is coupled with the dark matter detector to remove krypton or to remove radon by revised operation. The present paper mainly focuses on the offline operating mode. The flow diagram of the system is shown in Fig.~\ref{fig3}. The offline mode mainly includes the original xenon feeding line, the off-gas discharging line, and the product xenon discharging line, which are colored in red, green and magenta, respectively.

For the feeding line, the original xenon in the gas cylinder flows through the filter to remove particulate impurities above 0.5 microns firstly. Then it goes through a reduction valve and a flow controller (FM2), which are able to decrease the gas pressure to about 250 kPa and control the flow rate at 30 standard liters per minute (slpm). After that the gas flows through a getter to reduce the concentration of O$_2$, N$_2$, CO$_2$, CO, H$_2$, CH$_4$ and H$_2$O to below 1$\times10^{-9}$ mol/mol and enters the distillation apparatus. The original gas xenon is pre-cooled by the product liquid xenon that flows out from the reboiler at a heat exchanger. Then the refrigerator (Model PT60, Cryomech. Inc), installed downstream of the heat exchanger, is responsible for further condensing the original xenon gas into 178 K liquid. At last, the saturated liquid xenon flows into the distillation tower through the feeding point 1 in Fig.~\ref{fig3}, and starts heat and mass transfer process at the packings of the distillation column. The cryogenic distillation column is thermal insulated by high vacuum multilayer insulation, and the vacuum is maintained at about 3$\times10^{-3}$ Pa by a vacuum pump.

The discharging line is simpler compared with the feeding line. The off-gas is collected to an aluminum alloy bottle from the top condenser at a flow rate of 0.3 slpm (0.1 kg/h). For this purpose, the flow controller (FM1) is adopted and the collection bottle is put into a dewar filled with liquid nitrogen. The product collecting line, starting from the bottom of the reboiler, is similar to the off-gas discharging line. The differences are the product liquid xenon exchanges heat in the heat exchanger to the original gas xenon before flowing out of the system and the flow controller (FM3) is set to 29.7 slpm (9.9 kg/h).

In the online krypton removal mode, which is colored in blue as shown in Fig.~\ref{fig3}, the system is coupled with dark matter detector. The product xenon flows directly to the dark matter detector via a pump at a flow rate controlled by the flow controller (FM4), then the circulating pump feeds the xenon back into the distillation system. The other unnoted pipelines in Fig.~\ref{fig3} are for the purpose of online radon removal and sampling.

\subsection{The operation procedure}

The operation procedure of the distillation system is divided into 5 processes: pre-cooling process, gas charging process, total reflux process, purification process and collection process.

The pre-cooling process is the first operating process after start-up. The system is charged with 250 kPa xenon gas at room temperature, and the refrigerator (Model AL300 in Fig.~\ref{fig3}, Cryomech. Inc) is switched on to cool the inner tower and liquify the original xenon gas at the working temperature of 178 K. During this process, it is necessary to charge the original xenon into the tower, in order to avoid pressure decreasing along with cooling and maintain the pressure of the distillation tower at about 200 kPa, so as to make sure the xenon can be liquefied at 178 K. When liquid xenon appears in the reboiler at the bottom of the distillation tower, it means the structured packings have been wetted and the process moves on.

In the gas charging process, the original xenon gas is continuously recharged into the distillation tower at a flow rate of ~15 slpm until the liquid level in the reboiler reaches 15 cm.

In the total reflux process, the original xenon gas feeding is stopped, and the heating power of 120 W is added to the reboiler. After multiple evaporation and condensation, a gas-liquid equilibrium state is established in the distillation tower, and the xenon is redistributed in two phases regarding of the krypton concentration. The liquid product xenon with less krypton is in the reboiler, and the off-gas enriched with krypton is in the condenser.

The purification process is a dynamic distillation process. The original xenon is charged into the system at a flow rate of 30 slpm, while the product xenon and off-gas flow out of the system at the flow rate of 29.7 slpm and 0.3 slpm, respectively. During this process, the product xenon is sampled and measured to monitor if the krypton concentration is qualified.

In the collection process, the original xenon feeding is stopped and the system is shut down after all the xenon has been collected. The product xenon and off-gas are collected according to the ratio of 100:1. When the liquid level in the reboiler reduces to 0.1 cm, in order to ensure the purity of the product xenon, the product xenon discharging process is ended, and the remaining xenon is collected as off-gas from the top of the tower.

\section{Operation data and analysis\label{sec4}}
\subsection{Pre-cooling process}

\begin{figure} 
    \centering 
    \includegraphics[width=0.45\textwidth]{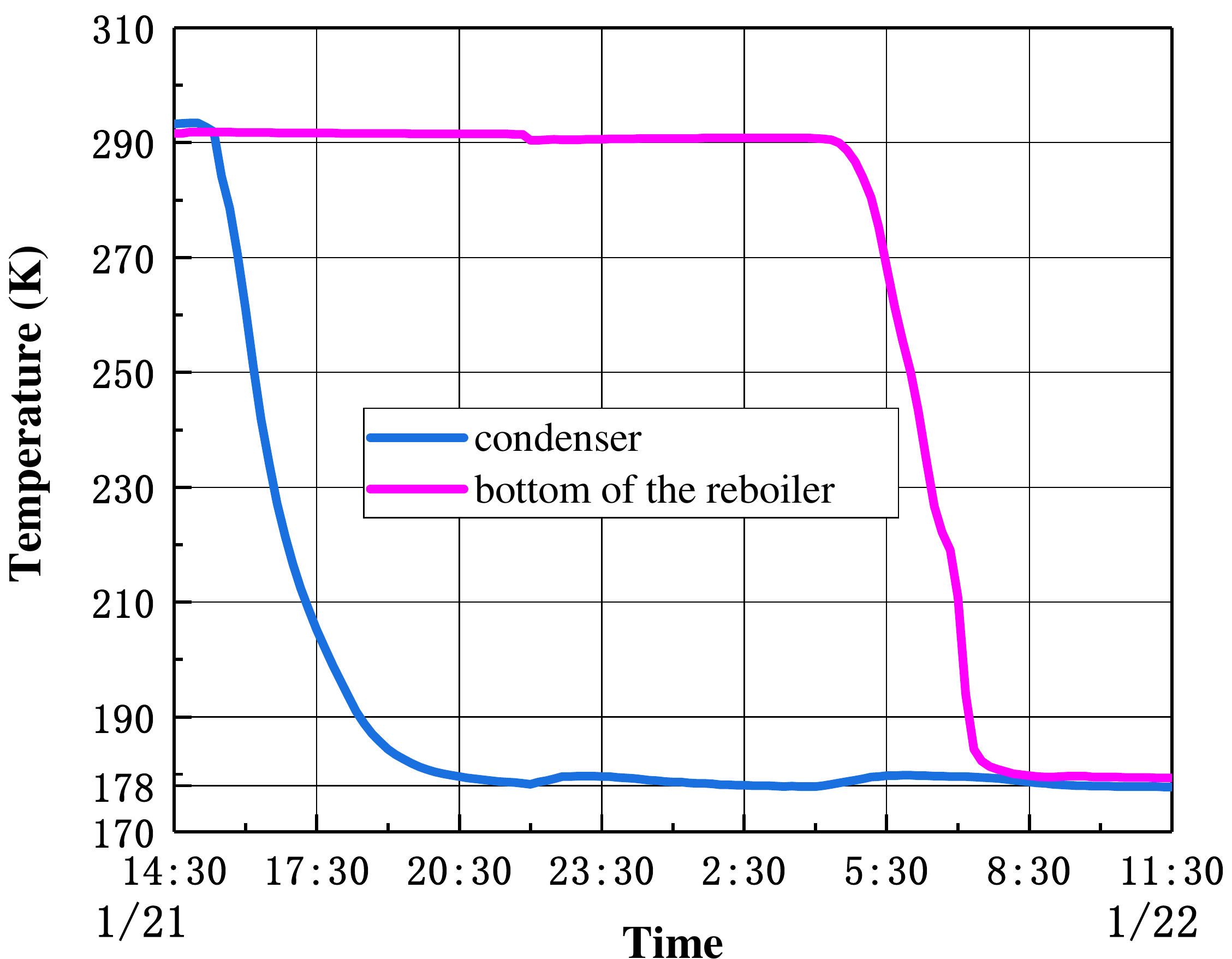}\\ 
    \caption{Temperature variations during the pre-cooling process.\label{fig4}}
\end{figure}

The temperature variations of the condenser and the reboiler at the pre-cooling process are presented in Fig.~\ref{fig4}. As can be seen, the temperature of the condenser rapidly decreased from 293 K to 179.5 K within 6 hours after start-up, and then it gradually stabilized to 178 K, which is the working temperature of the refrigerator (Model AL300 in Fig.~\ref{fig3}). At the 14th hour, the temperature at the bottom of the reboiler began to drop sharply, indicating that liquid xenon began to appear in the reboiler, and it reached to 179.5 K after 18 hours. In general, the cold sources of refrigerators AL300 and PT60 cooled the reboiler from room temperature to about 179.5 K within 18 hours, and the cooling efficiency was relatively high comparing to the previous design\cite{16}.

\subsection{Gas charging process}

\begin{figure}[htbp]
\centering
\subfigure[Temperature and pressure variations during the gas charging process.]{
\label{fig5(a)}
\begin{minipage}[t]{0.45\textwidth}
\centering
\includegraphics[width=6.5cm]{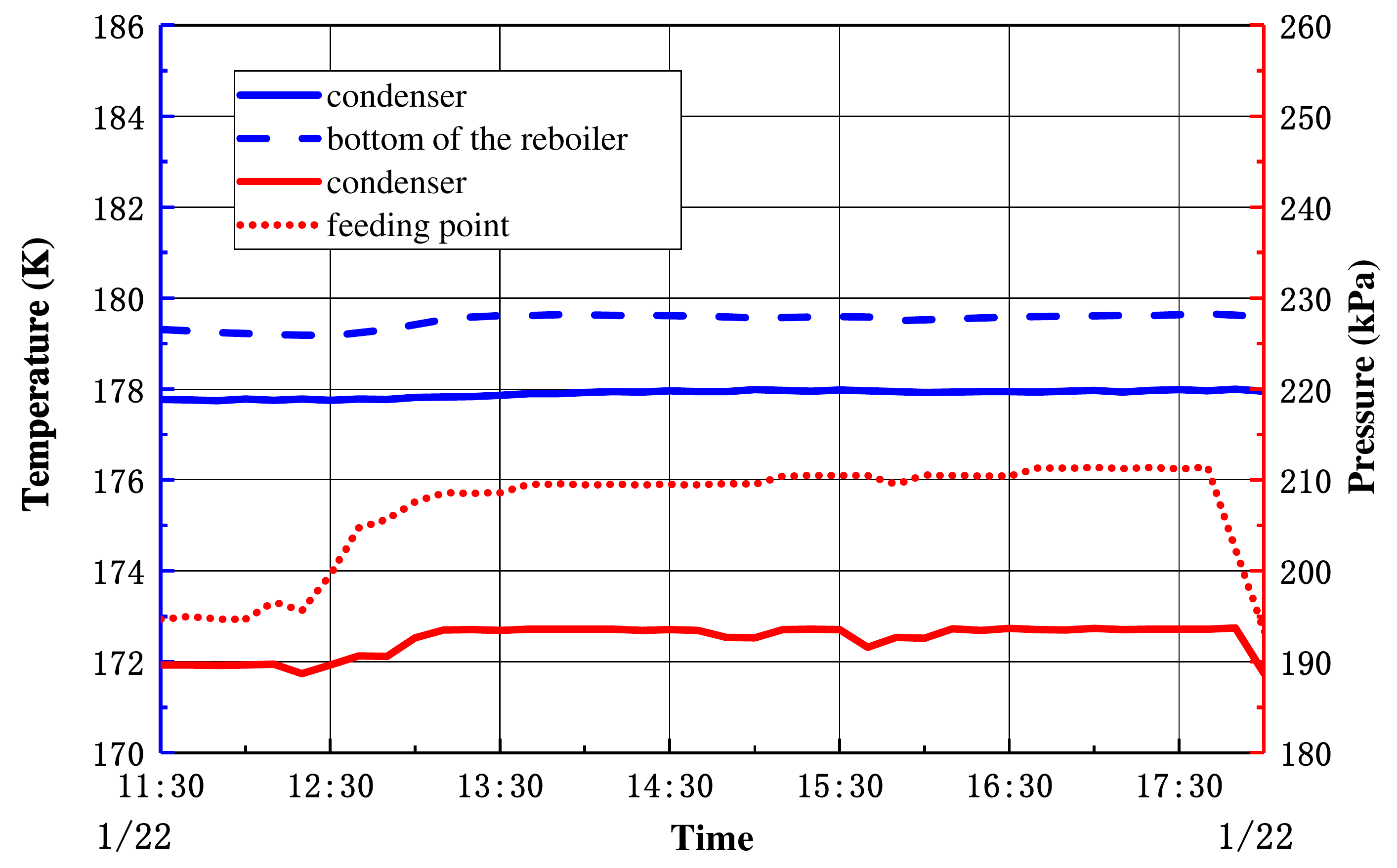}
\end{minipage}
}
\subfigure[Flow rate and liquid level variations during the gas charging process.]{
\label{fig5(b)}
\begin{minipage}[t]{0.45\textwidth}
\centering
\includegraphics[width=6.5cm]{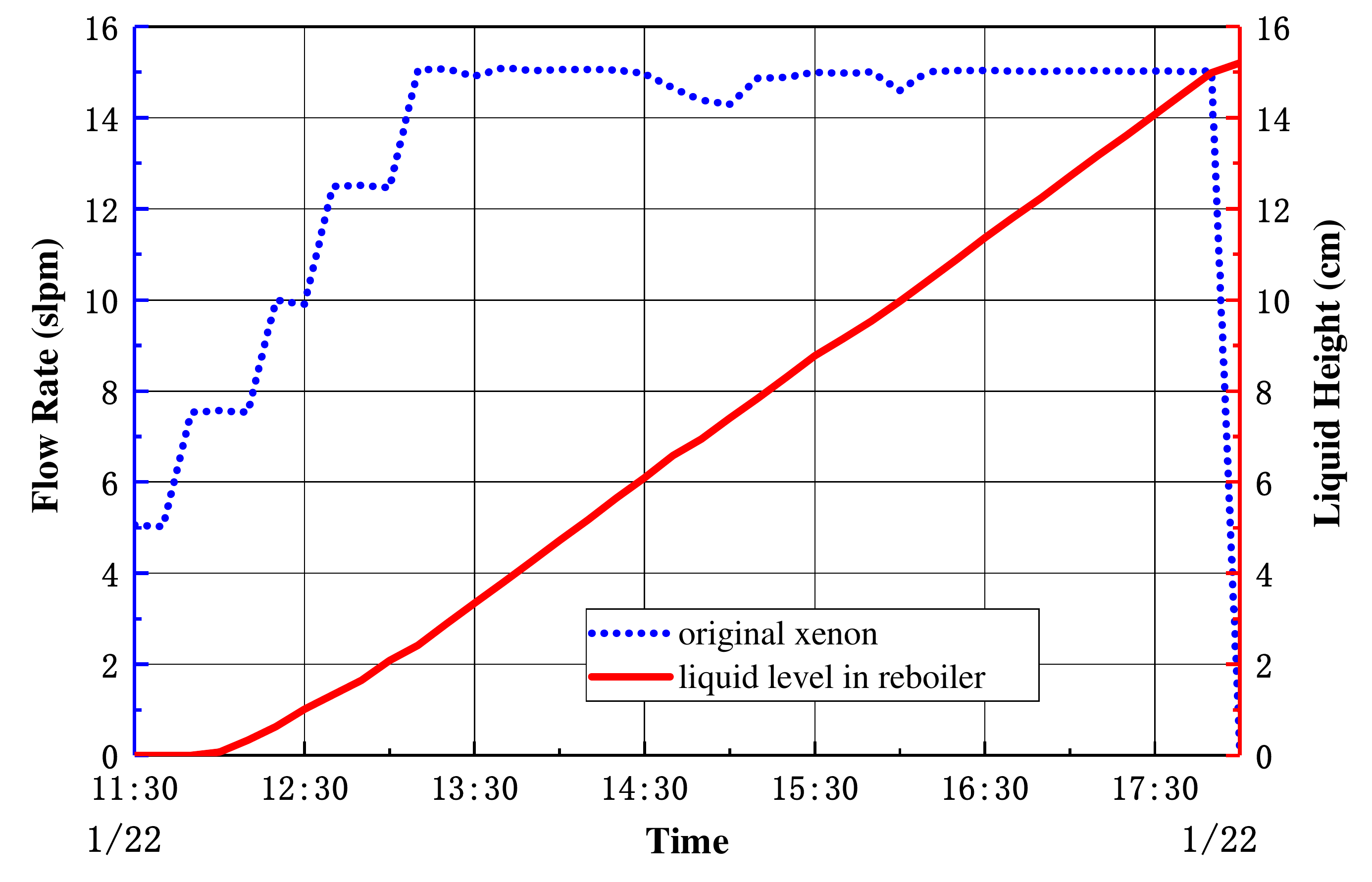}
\end{minipage}
}
\caption{ \label{fig5}Thermodynamic performance of the distillation system during the gas charging process.}
\end{figure}

At the gas charging stage, the height of the liquid level in the reboiler needs to reach the target value of 15 cm. The changes of related parameters are presented in Fig.~\ref{fig5}. Fig.~\ref{fig5(a)} illustrates the temperature and pressure variations of the distillation system. The feeding pressure was 210 kPa, which was higher than the system pressure in order to push the xenon into the distillation tower, and the pressure of the condenser increased from 190 kPa to 194 kPa slightly due to the feeding. The temperatures of the reboiler and the condenser remained at 179.5 K and 178 K, respectively, which means the feeding did not result in temperature vibration because the liquid xenon existed in the distillation tower and the saturated condition was achieved. As to the subgraph Fig.~\ref{fig5(b)}, it illustrates the flow rate change of the feeding xenon gas and the liquid level variation inside the reboiler. The feeding flow rate undergone multiple adjustments before it reached to 15 slpm in the first 1.5 hours. The feeding stopped when the liquid height reached to the aimed value of 15 cm after 27.5 hours.

\subsection{Total reflux process} 

\begin{figure} 
    \centering 
    \includegraphics[width=0.45\textwidth]{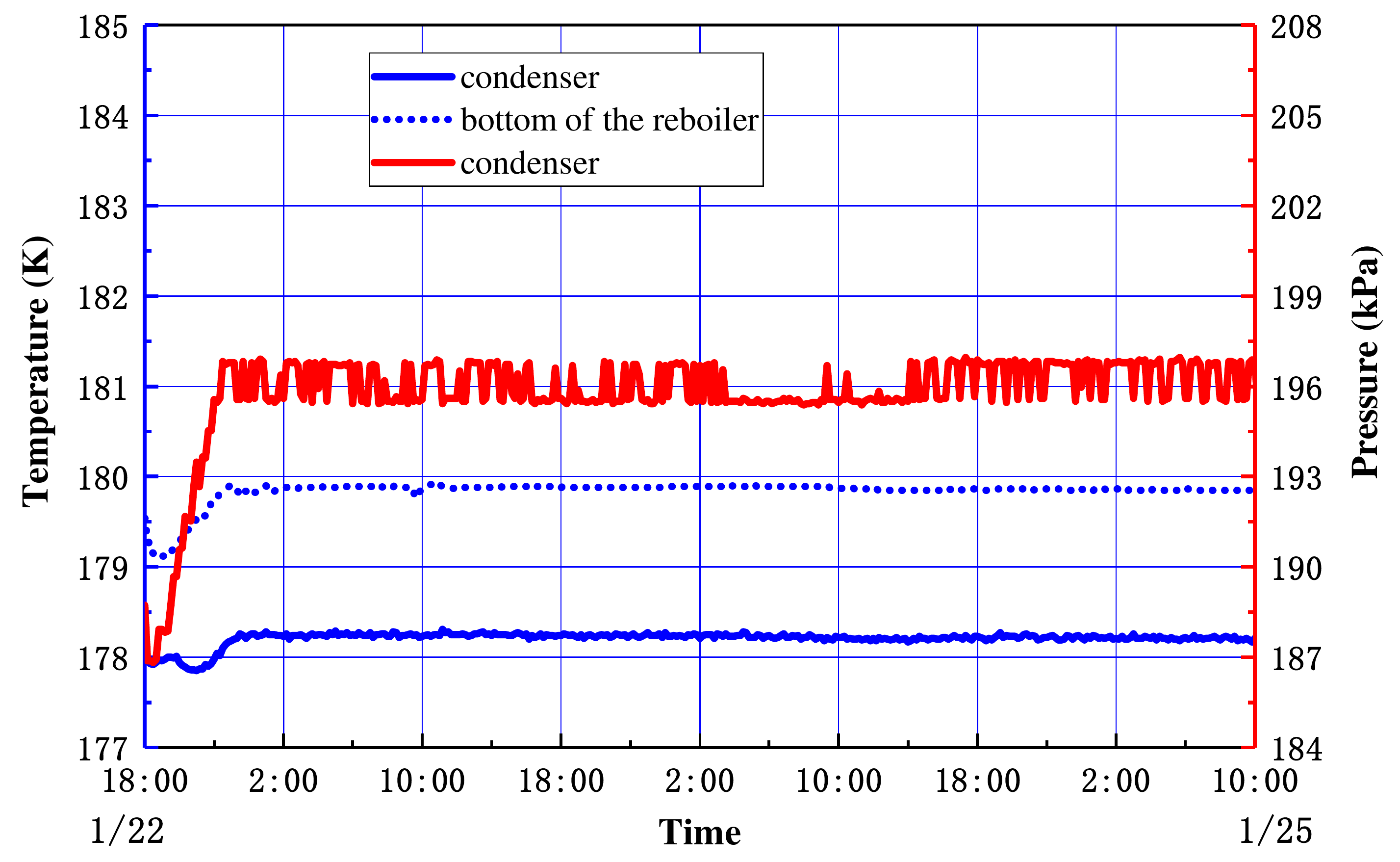}\\ 
    \caption{Temperature and pressure variations during the total reflux process. \label{fig6}}
\end{figure}

In the total reflux process, the feeding of original xenon was stopped and the heating power in the reboiler was added till 120 W. The xenon in the reboiler evaporated by heating and established an equilibrium with the refluxed liquid xenon falling from the condenser at the top of the tower. From January 22nd to January 25th, the thermodynamic stability of the distillation was set up in this process which can be seen in Fig.~\ref{fig6}. The temperatures of the reboiler and condenser were kept at 179.9 K and 178.2 K, respectively. The pressure in the condenser increased and fluctuated slightly around 196 kPa after the feeding stopped and the heating power of 120 W was added to the reboiler.

\subsection{Purification process}

\begin{figure}[htbp]
\centering
\subfigure[Temperature and pressure variations during the purification process.]{
\label{fig7(a)}
\begin{minipage}[t]{0.45\textwidth}
\centering
\includegraphics[width=6.5cm]{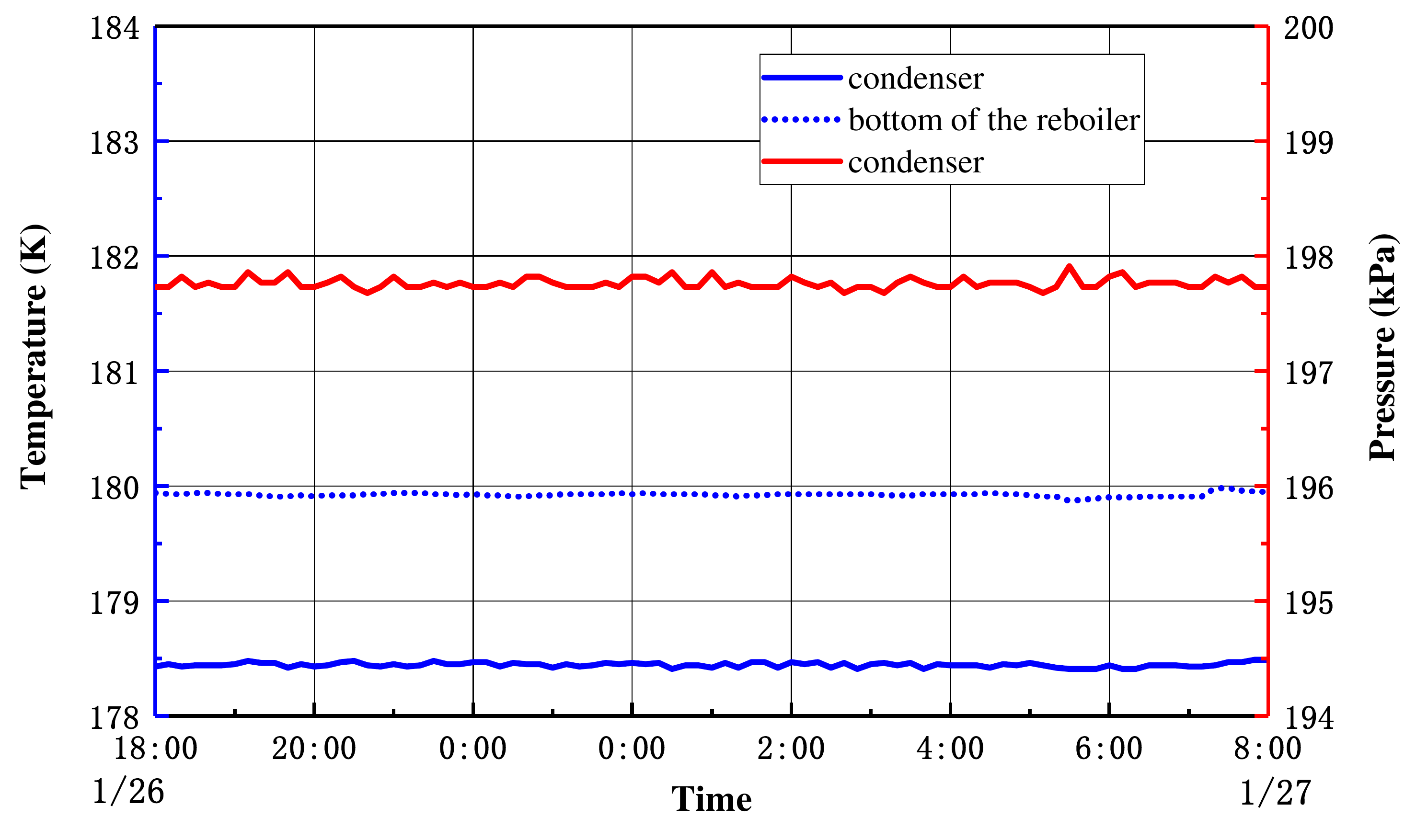}
\end{minipage}
}
\subfigure[Flow rate and liquid level variations during the purification process.]{
\label{fig7(b)}
\begin{minipage}[t]{0.45\textwidth}
\centering
\includegraphics[width=6.5cm]{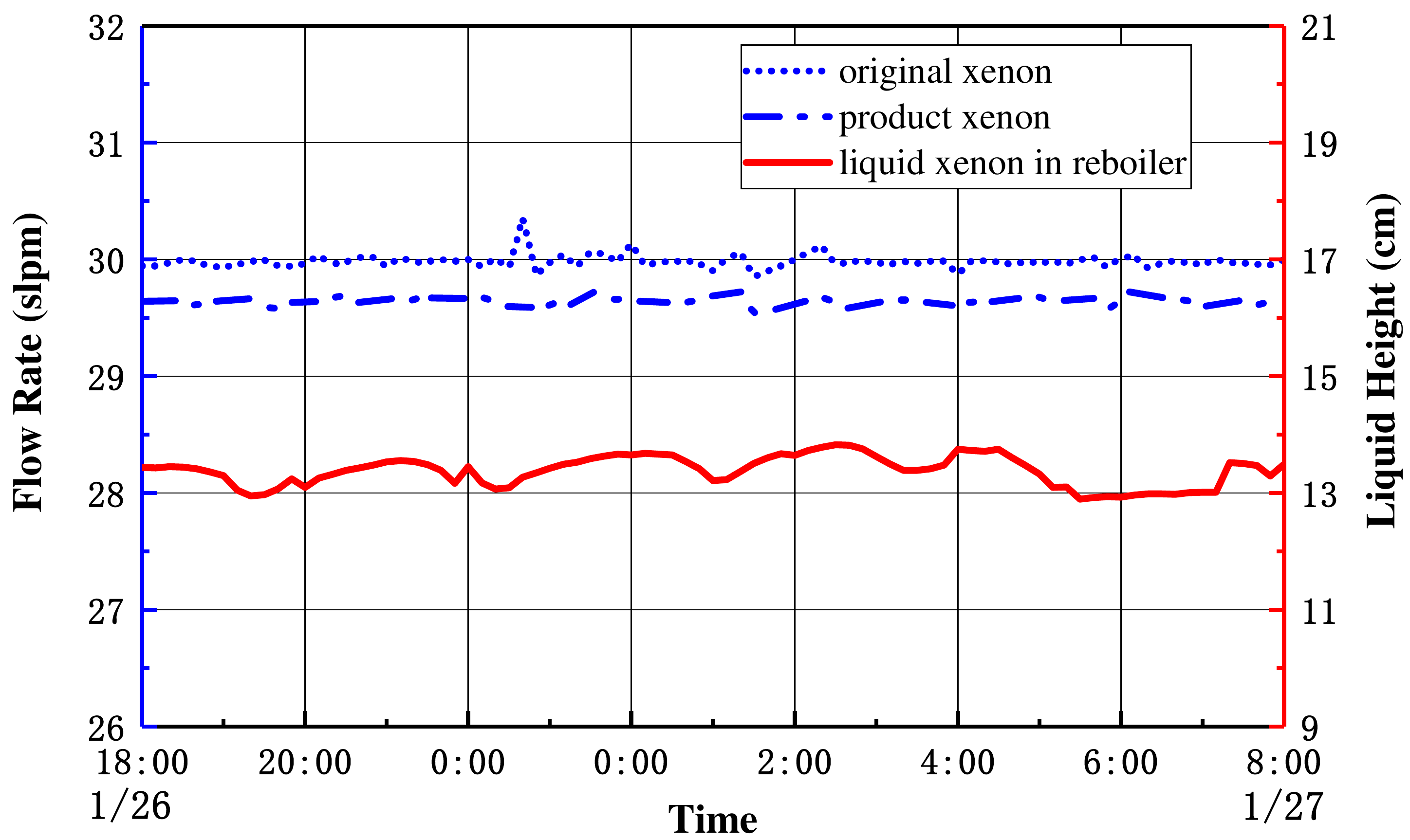}
\end{minipage}
}
\caption{ \label{fig7}Thermodynamic performance of the distillation system during the purification process.}
\end{figure}

The purification process lasted 40 days in total, a short period during the purification process which is characteristic to reflect important parameters of the thermodynamic stability in the purification process are presented in Fig.~\ref{fig7}. The line charts in Fig.~\ref{fig7(a)} illustrate the temperature and pressure variations of the distillation tower. The temperature of the bottom reboiler and the top condenser were maintained at 179.9 K and 178.5 K, respectively, and the condenser pressure fluctuated slightly at 197.7 kPa. As to the subgraph (b), it shows the flow rate variations of the feeding and discharging, as well as the liquid level variation in the reboiler. Except for occasional fluctuations, the feeding rate of the original xenon was stable at 30 slpm (10 kg/h), and the discharging rate of the product xenon was stabilized at 29.7 slpm (9.9 kg/h), while the off-gas rate was 0.3 slpm (0.1 kg/h). In the meanwhile, the corresponding liquid level in the reboiler vessel remained at about 13.5 cm. The above data indicates the thermodynamic stabilization worked efficiently during the purification process.

\subsection{Collection process}

\begin{figure}[htbp]
\centering
\subfigure[Temperature variations during the collection process.]{
\label{fig8(a)}
\begin{minipage}[t]{0.45\textwidth}
\centering
\includegraphics[width=6.5cm]{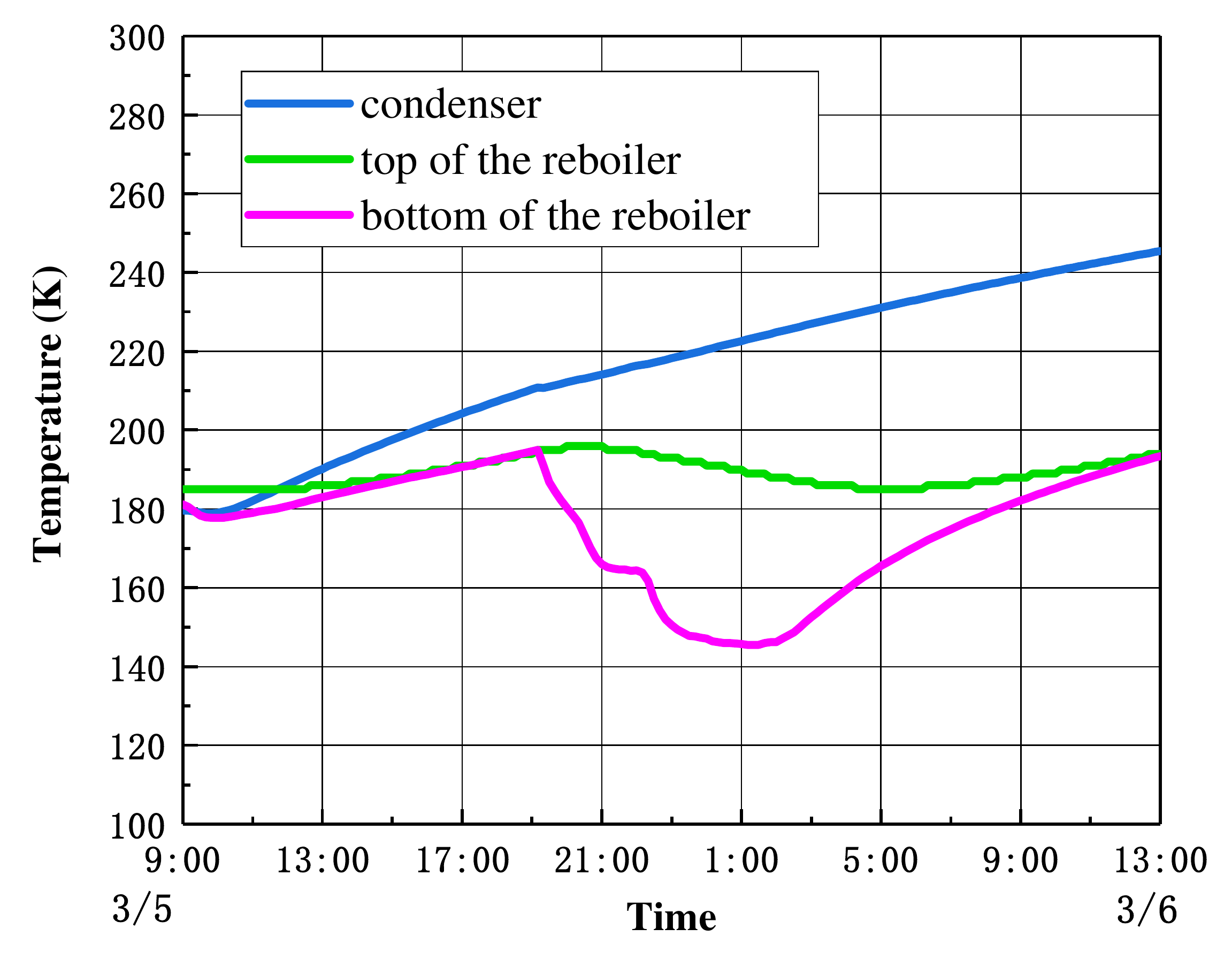}
\end{minipage}
}
\subfigure[ Pressure variations during the collection process.]{
\label{fig8(b)}
\begin{minipage}[t]{0.45\textwidth}
\centering
\includegraphics[width=6.5cm]{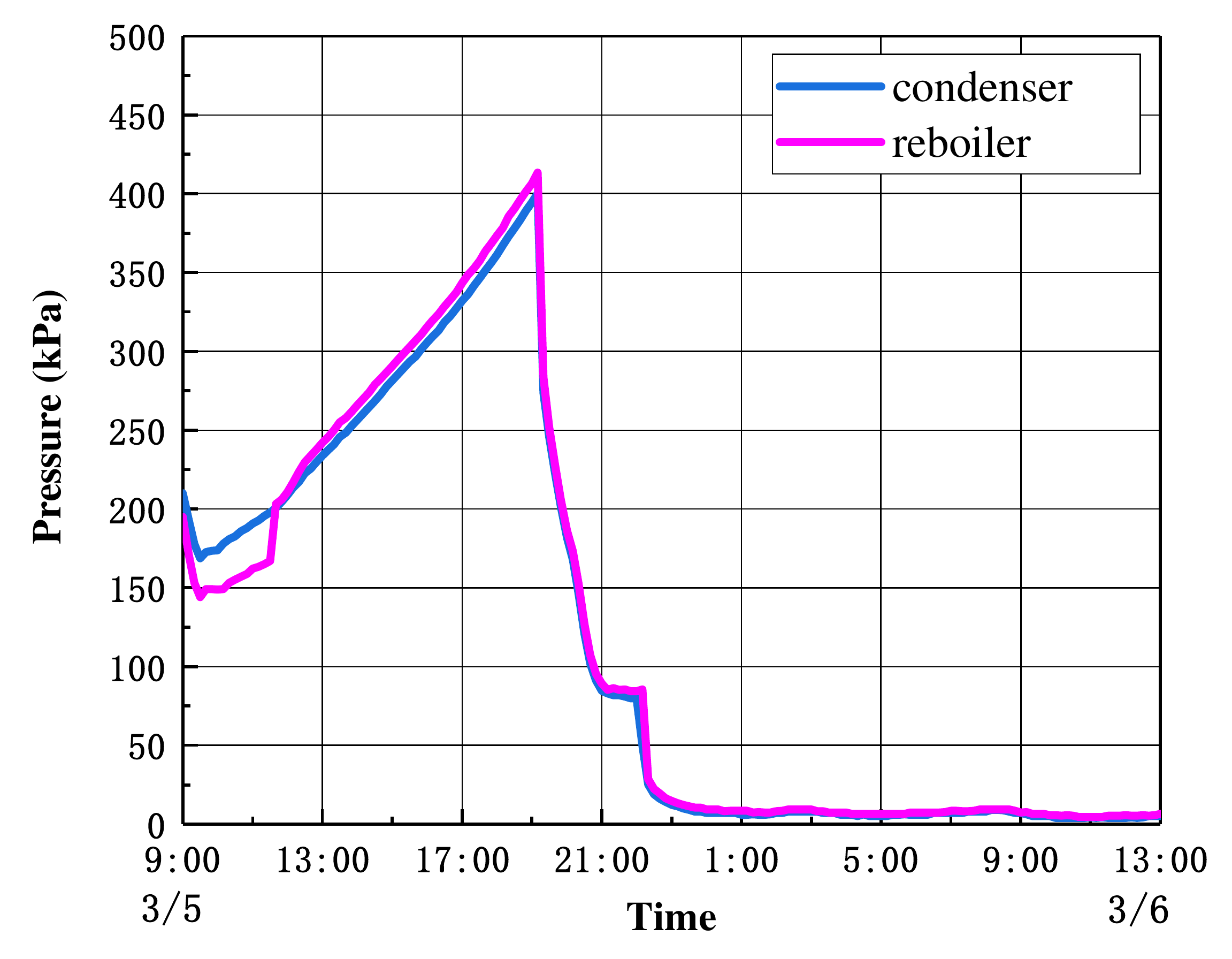}
\end{minipage}
}
\caption{ \label{fig8}Temperature and pressure variations of the distillation system during the collection process.}
\end{figure}

\begin{table}[htbp]
    \caption{Saturation parameter of xenon. \label{table2}}
    \begin{tabular}{ccc}
        \toprule
   Saturation state&Pressure (kPa)&Temperature (K)  \\
        \midrule
      &20&144.5\\
    Solid-vapor&40&152.1\\
    saturation&60&157.1\\
      &80&161.0\\
        \midrule
    &85&162.2\\
    &100&172.2\\
    &150&177.9\\
    Vapor-liquid&200&182.6\\
    saturation&250&186.6\\
    &300&190.2\\
    &350&193.4\\
    &400&196.3\\
        \toprule
    \end{tabular}
\end{table}

The last process is collection process. During this process, the feeding of original xenon was finished, and the product xenon and off-gas were collected according to the ratio of 100:1. The refrigerator was shut down, and the heating power in the reboiler was stopped when the liquid level was less than 7 cm for the height of the heating rods to avoid damage. The temperature and pressure curves are shown in Fig.~\ref{fig8}. In the first 10 hours, the heat leakage was the key factor to influence the temperature and pressure, it caused the temperature of the system rising slowly and the liquid xenon in the reboiler evaporating constantly. Because the gas collection rate (the off-gas collection rate) was slower than the evaporation rate, the pressure of the tower kept increasing at the initial 10 hours. At 19:00 on March 5th, the temperature at the bottom of the reboiler was the same as that at the top, indicating that there was almost no liquid xenon in the reboiler, so the discharge became the vital factor to influence the pressure and caused the tower pressure to drop sharply. With reference to the data in Table~\ref{table2}, it can be found that the xenon in the reboiler was basically saturated, and when the saturation pressure dropped, the corresponding saturation temperature also decreased. Therefore, between 19:00 on March 5 and 1:00 on March 6, the temperature in the reboiler decreased due to the pressure dropping. After 1:00 on March 6, the temperature in the reboiler rose up again because of the heat leakage when the liquid xenon in the reboiler was almost collected. The xenon gas in the condenser was at overheated condition according to Table~\ref{table2}, so its temperature kept increasing as the result of heat leakage.

\section{Kr concentration of distillated product Xe \label{sec5}}

\begin{figure*}[htbp]
\centering 
\includegraphics[width=\textwidth]{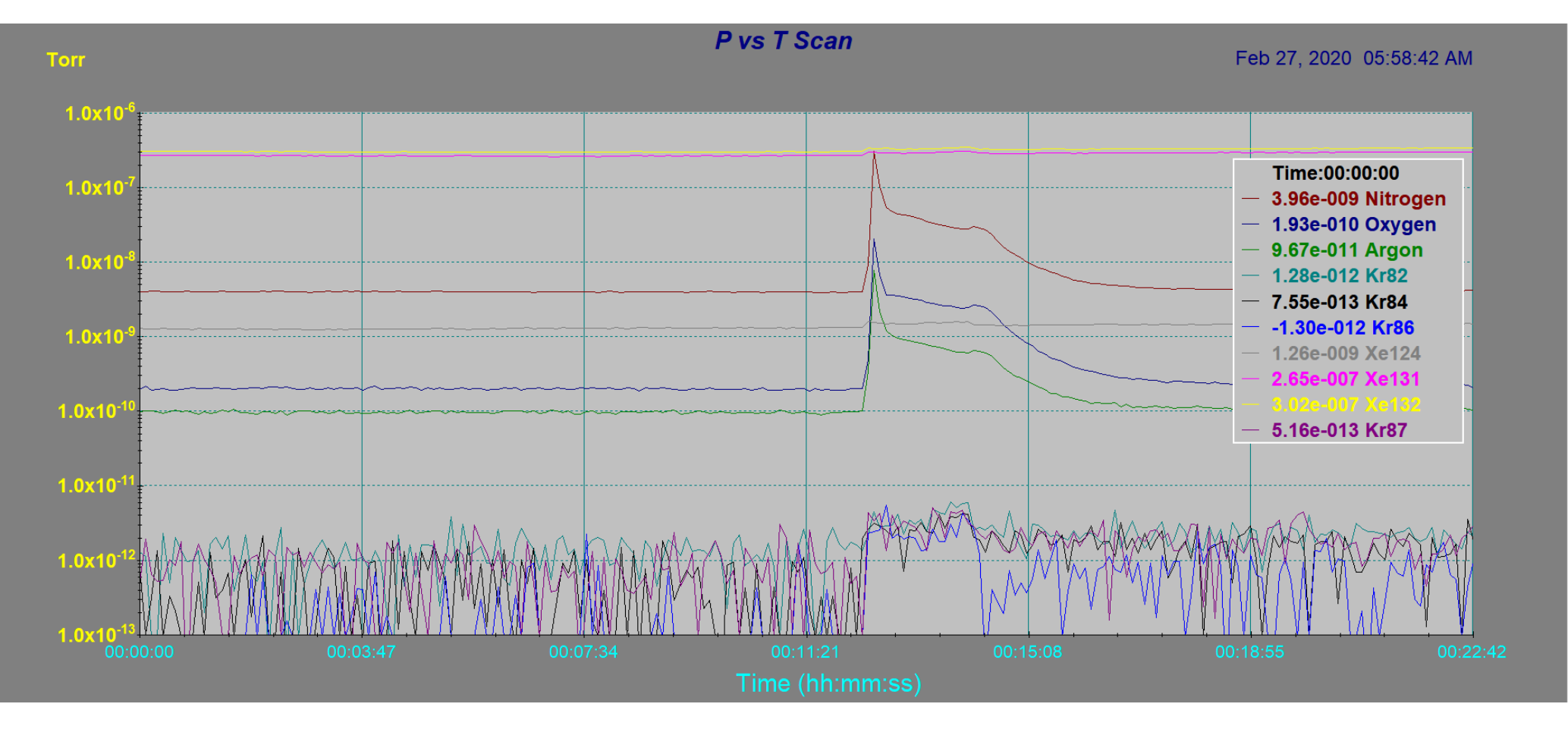}
\caption{The RGA spectrum of the purified xenon measuring result. \label{fig9}}
\end{figure*}

In order to measure the krypton concentration in the product xenon, a system has been developed and calibrated by xenon samples with known krypton concentration, which allows detecting krypton impurities at ppt level in principle. The measurement and calibration methods are illustrated in Ref.~[21]. The krypton measurement and analysis system are based on a residual gas analyzer (RGA) and a liquid nitrogen cold trap. One of the key factors to measure the krypton concentration as low as ppt level is the relative enrichment of krypton in xenon, which is achieved by the liquid nitrogen cold trap. The krypton concentration in product xenon can be obtained by analyzing the partial pressure of different component species measured by the RGA set behind the liquid nitrogen cold trap. More details of the krypton measurement system are given in Ref.~[22]. Though the krypton concentration of the distilled xenon is at ppq level by design, only an upper limit can be given for the sensitivity of this krypton measurement system.

Fig.~\ref{fig9} presents the RGA spectrum of the product xenon of the distillation system. The gas partial pressure ratio of $^{84}$Kr to $^{132}$Xe is a very critical parameter, which reflects the krypton content. The krypton concentration can be calculated by comparing the parameter $^{84}$Kr/$^{132}$Xe between the product xenon and the xenon sample. Based on a series of analysis and calculations with the monitored data in Fig.~\ref{fig9}, the krypton concentration in the product xenon was measured to be <8.0 ppt (@90$\%$ C. L).

\section{Conclusion\label{sec6}}

A large scale cryogenic distillation system for removing krypton from xenon has been designed, constructed and operated for reducing the krypton concentration from 5$\times10^{-7}$ mol/mol to $\sim10^{-14}$ mol/mol theoretically with a purification speed of 10 kg/h and the collection efficiency of 99$\%$. The technological process, operation procedure and experimental data analysis of the distillation system are presented in this paper. The distillation system has been working stably for 1.5 months and produced 5.75 tons of ultra-high purity xenon for the PandaX-4T dark matter detector. Thermodynamic analysis shows that the distillation column performs steady during the purification process. The upper limit of the krypton concentration in product xenon was measured to be 8.0 ppt. The experimental analysis of PandaX-4T ultra-high purity cryogenic distillation system is also an important reference for optimizing the distillation operation.

\acknowledgments
The authors would like to thank the supports of the PandaX-4T collaboration. 
This project is supported by grants from the Ministry of Science and Technology of China (No. 2016YFA0400301 and 2016YFA0400302), a Double Top-class grant from Shanghai Jiao Tong University, grants from National Science Foundation of China (Nos. 11435008, 11505112, 11525522, 11775142 and 11755001), grants from the Office of Science and Technology, Shanghai Municipal Government (Nos. 11DZ2260700, 16DZ2260200, and 18JC1410200), and the support from the Key Laboratory for Particle Physics, Astrophysics and Cosmology, Ministry of Education. This project is supported by Sichuan Science and Technology Program (No.2020YFSY0057). 
We also thank the sponsorship from the Chinese Academy of Sciences Center for Excellence in Particle Physics (CCEPP), Hongwen Foundation in Hong Kong, and Tencent Foundation in China. Finally, we thank the CJPL administration and the Yalong River Hydropower Development Company Ltd. for indispensable logistical support and other help.

\section*{Data Availability Statement}

The data that support the findings of this study are available from the corresponding author upon reasonable request.

\end{document}